\documentclass[letterpaper, 10 pt, conference]{ieeeconf}  

\usepackage{graphicx}      

\usepackage{latexsym}
\usepackage{amsmath}
\usepackage{amsfonts}
\usepackage{epstopdf}
\usepackage{amssymb}

\usepackage{float}
\usepackage{soul}
\usepackage{xcolor}
\usepackage{comment}
\newtheorem{example}{Example}
\newtheorem{definition}{Definition}

\newtheorem{theorem}{Theorem}
\newtheorem{lemma}[theorem]{Lemma}

\usepackage{color}
 
\usepackage{xcolor}
\definecolor{verde}{rgb}{0,0.7,0}

\usepackage[colorlinks=true,linkcolor=blue,urlcolor=blue,citecolor=blue]{hyperref}

\newcommand{\be}{\begin{equation}}
\newcommand{\ee}{\end{equation}}

\newcommand{\bea}{\begin{eqnarray}}
\newcommand{\eea}{\end{eqnarray}}
\newcommand{\bean}{\begin{eqnarray*}}
\newcommand{\eean}{\end{eqnarray*}}

\IEEEoverridecommandlockouts                              

\begin{document}

\title{\LARGE \bf
Tripartite and Sign Consensus for Clustering Balanced Social Networks}

 \author{Giulia De Pasquale and Maria Elena Valcher 
 \thanks{G. De Pasquale and M.E. Valcher are with
 the Dipartimento di Ingegneria dell'Informazione
 Universit\`a di Padova, 
    via Gradenigo 6B, 35131 Padova, Italy, e-mail:  \texttt{giulia.depasquale@phd.unipd.it, meme@dei.unipd.it}.
This Paper is an extended version of the paper G. De Pasquale, M.E. Valcher. \textit{Tripartite and Sign Consensus for Clustering Balanced Social Networks}. Proceedings of the ACC 2021, New Orleans, LA, USA.
}
   } 
 \maketitle

\begin{abstract}                          
In this paper we address two forms of consensus for  multi-agent systems with  undirected, signed, weighted, and  connected communication graphs, under the assumption that the agents can be partitioned into three clusters, representing the decision classes on a given specific topic, for instance, the in favour, abstained and opponent agents. 

We will show that under some     assumptions on  the cooperative/antagonistic relationships   among the agents,
simple modifications of   DeGroot's algorithm allow to achieve tripartite consensus
 (if the opinions of agents belonging to the same class all converge to the same decision) or sign consensus (if the opinions of the agents in the three clusters converge to positive, zero and negative values, respectively).
 \end{abstract}
 
%

\section{Introduction} 
Social networks consisting of a finite number of individuals that mutually interact in a cooperative or antagonistic way are frequently represented by signed graphs,
whose positive edges represent friendly ties, while negative edges 
denote  enmity relationships.  
Generally speaking, consensus for a network of agents is the problem of achieving a common objective, of converging to a common decision, by making use of information provided by neighbouring agents. 
 During the past few decades, consensus problems have attracted the attention of scientists and researchers from various fields such as sociology, engineering and mathematics, as it can be seen from the large amount of scientific literature related to this topic \cite{Altafini2013,Bullo2020,davis67,DeGroot,Heider}. However, these problems have been  typically investigated under the assumption  that the overall communication network is  purely cooperative (pure ``consensus")  \cite{OlfatiFaxMurray,OF-Murray2004,RenBeardAtkins} or ``structurally balanced", by this meaning that agents split into two groups of cooperative agents that compete with those of the other group (``bipartite consensus") \cite{Altafini2013,Bauso1,Easley,ValcherMisra}.

On the other hand, when focusing on  signed graphs that represent meaningful social relationships, 
cooperation and structural balance are only two of the possible  sign configurations, and 
additional models have been considered 
\cite{Johnsen}.  If we restrict our attention to the case when the interpersonal appraisals between agents are reciprocal, namely the signed graph that models the interactions between the agents is undirected, in addition to   structural balance also   ``clustering balance" may arise \cite{Bullo2020}.  Specifically, when the sign attribution over the graph network is given according to the following rules: 1) the friend of my friend is my friend, 2) the enemy of my friend is my enemy, 3) the friend of my enemy is my enemy, clustering balance is obtained. If in addition to the previous three rules, the rule: 4) the enemy of my enemy is my friend is observed, then the network is structurally balanced. The aforementioned four rules are  a cornerstone of the research on the opinion dynamics in social networks and are known in literature as "Heider's   rules" \cite{Heider}.

To the best of our knowledge,  the literature on  consensus problems over clustering balanced networks, by this meaning the problem of making all agents belonging to the same cluster in a clustering balanced network achieve a common decision,   is quite limited \cite{GiuliaElenaAut2020, GiuliaElenaCDC2020}.
On the other hand, there have been some interesting research efforts aiming to explore the possibility of achieving {\em group consensus} 
for networks whose agents have been partitioned in disjoint groups. Such group partition, however, does not represent a balanced clusterization, according to Heider's rules,   and hence it is
not suitable for formalising  consensus problems  in a sociological  context \cite{QinMaZhengGao,QinYu2013,QinYuAnderson2016,XiaCao2011,YuWang2010}. Indeed, in all of these papers, the group partitioning is obtained 
according to the ``indegree balanced condition", that ensures that agents within the same group cooperate, while each agent has both cooperative and antagonistic relationships with the agents of every other group, but the weights of such relationships sum up to zero.

The ambition of this paper is to fill in a gap between the scientific literature regarding clustering balanced networks  and the one related to consensus, by proposing two forms of  consensus problems on    clustering balanced networks and by providing conditions for their solvability. Specifically, given an undirected, signed, weighted, and connected network, with three disjoint and antagonistic clusters,  we investigate under what conditions the opinions of  (cooperative) agents belonging to the same cluster   converge to the same value/decision
or at least they  converge to 
values/decisions having the same sign, and such a sign varies with the specific cluster
(so, one cluster converges to a positive decision, one to a  negative one and the members of the third cluster all converge to the zero value).
These two targets correspond to two different notions of consensus that we will refer to as {\em tripartite consensus} and  {\em sign consensus}, respectively. 
The sociological interpretation of these two problems 
is easily found  in  contexts such as elections, group decisions, bets,  and every time   agents are called to express their approval, disapproval, or abstention on a given   topic or decision (see  \cite{Jiang}) and hence   split into three classes.\\
{\color{black} Additional applications of these problems can be found in rendezvous  problems for multi-robots systems or formation flights.}\\
The results  presented in this paper have been inspired by the work of C. Altafini \cite{Altafini2013}, where  the concept of {\em bipartite consensus} has been introduced for structurally balanced networks, by the tutorial paper of A.V. Proskurnikov and R. Tempo \cite{ProskurnikovTempo} on social networks, and by the works of J. Davis  \cite{davis67} and P. Cisneros-Velarde and F. Bullo \cite{Bullo2020}, where the concept of clustering balance plays a major role.\\
In our recent papers \cite{GiuliaElenaAut2020,GiuliaElenaCDC2020} we have investigated consensus
problems   for clustering balanced networks\footnote{From now onward when referring to clusters of a communication network/signed graph we will always assume that the network is clustering balanced and omit this specification.}
 with $3$  or, in general,   $k\ge 3$    clusters,
   under some ``homogeneity" constraint on the (positive and negative) weights  of the communication network, namely by assuming that the amount of trust/mistrust that each agent attributes to its friends/enemies is  prefixed for all the agents in the same cluster. In
   this paper we 
   will focus on networks that are partitioned into three clusters, and we will  first show that, even without the homogeneity assumption, (tripartite) consensus can be obtained by means of   a slightly revised version of   De Groot's distributed feedback control law. Subsequently, we will introduce     sign consensus  and show that also in that case, under some mild assumptions,  a modified version of De Groot's  control law allows to successfully achieve the target.

The   paper is organized as follows.   Notation and preliminaries are first   introduced. Section \ref{2} formalises the tripartite consensus problem for a multi-agent network, whose agents are described as simple integrators and whose communication  graph  splits into 3 clusters. Section \ref{3}  
provides a complete solution to this problem, under some  mathematical assumptions 
formalising the existence of a strong relationship among the individuals of at least one cluster and, on the contrary,   strong competition among the agents of adverse clusters. Section \ref{4} explores, under the same assumptions, the 
more general target of
sign consensus, namely the 
case when the three clusters asymptotically converge to a positive, negative and neutral (namely zero) decision, respectively, but 
this decision within each cluster is not necessarily of the same modulus, just of the same sign.

\smallskip

 {\bf Preliminaries}.\
For   $k, n\in \mathbb{Z}$,   $k \le n$,  we denote by   $[k,n]$    the  integer set  $\{k, k+1, \dots, n\}$.
The  symbol  $[A]_{i,j}$ denotes the $(i,j)$th entry of the matrix $A$, while  $[{\bf v}]_i$ is the $i$th entry of the vector ${\bf v}$. 
A matrix  (in particular, a vector)
 $A$ is   {\em nonnegative}  (denoted by  $A \ge 0$) \cite{BookFarina}  if all its entries are nonnegative.
 $A$ is   {\em strictly positive} (denoted by  $A \gg 0$) if all its entries   are positive.
A symmetric matrix $P\in {\mathbb R}^{n\times n}$  is {\em positive (semi) definite} if ${\bf x}^\top P {\bf x} >0$ (${\bf x}^\top P {\bf x} \ge 0$) for every ${\bf x}\in {\mathbb R}^n, {\bf x}\ne0,$ and when so we use the symbol $P\succ 0$ ($P\succeq 0$).\\
 The notation ${\mathcal A}= \text{diag}\{{\mathcal A}_1, \dots, {\mathcal A}_n\}$ indicates a block diagonal matrix with diagonal blocks  ${\mathcal A}_1, \dots, {\mathcal A}_n$. 
${\bf 0}_n$ and  ${\bf 1}_{n}$ are the 
   $n$-dimensional vectors  with all entries equal to $0$ and to $1$, respectively. 
%
A  real square matrix $A$ is 
{\em Hurwitz} if   every eigenvalue $\lambda$ in  $\sigma(A)$, the spectrum of $A$, has negative real part, i.e., ${\rm Re}(\lambda) <0$.
\\
A {\em Metzler matrix} is a real square matrix, whose off-diagonal entries are nonnegative.
For  $n\ge 2$,  an $n \times n$ nonzero Metzler  matrix $A$ is
 {\em reducible} \cite{F12b,Minc} if 
 there exists   a  permutation matrix $\Pi$ such that 
$\Pi^{\top} A \Pi$ is block-triangular, otherwise it is {\em
irreducible}.  
\\
Every  Metzler matrix $A$   
 exhibits a real dominant   (but not necessarily simple) eigenvalue  \cite{SonHinrichsen}, known as {\em Frobenius eigenvalue} and denoted
 by   $\lambda_{F}(A)$.  In other words, $\lambda_{F}(A) > {\rm Re}(\lambda), \forall\ \lambda \in \sigma(A), \lambda \ne \lambda_{F}(A)$. If $A$ is also irreducible, then  $\lambda_{F}(A)$ is necessarily simple.

The following technical result will be used extensively in this paper.

\begin{lemma}\label{lemma3} \cite{GiuliaElenaAut2020} Let $D\in {\mathbb R}^{n\times n}$ be a diagonal matrix and let $A\in {\mathbb R}^{n \times n}$ be a symmetric nonnegative matrix, then:
\begin{itemize}
\item[i)]  $D-A$ is positive definite if and only if there exists a strictly positive vector ${\bf v}\in {\mathbb R}^n$ such that $(D-A){\bf v} \gg 0$.
\item[ii)] If condition i) holds, then $(D-A)^{-1} \ge 0$ and is symmetric.
\end{itemize}
\end{lemma}
\smallskip

\section{Tripartite consensus: Problem statement}\label{2}

We consider a multi-agent system consisting of $N$ agents, each of them described as a continuous-time integrator  (see \cite{Altafini2013,OlfatiFaxMurray,OF-Murray2004,RenBeardAtkins}). The overall system dynamics is described as
\be 
\dot{\bf x}(t) = {\bf u}(t), 
\label{model}
\ee
where ${\bf x}\in \mathbb{R}^{N}, {\bf u}\in \mathbb{R}^{N}$, are the state and input variables, respectively.
\\
{\bf Assumption 1 on the communication structure.}\ [Connectedness and clustering]\ The communication among the $N$ agents is described by an   undirected, signed and  weighted communication graph ${\mathcal G}= ({\mathcal V}, {\mathcal E}, {\mathcal A})$, where ${\mathcal V}=\{1,2,\dots, N\}$ is the set of vertices, ${\mathcal E}\subseteq {\mathcal V} \times {\mathcal V}$ is the set of arcs, and ${\mathcal A}$ is the adjacency matrix of ${\mathcal G}$ that describes how agents interact. The 
 $(i,j)$th entry of ${\mathcal A}$, $[{\mathcal A}]_{i,j}$, $i\ne j$, is nonzero if and only if 
 the information about the status of the $j$th agent is available to the $i$th agent.   
We assume that the interactions between pairs of agents are symmetric and hence ${\mathcal A}={\mathcal A}^\top$.
The interaction between the $i$th and the $j$th agents is cooperative if $[{\mathcal A}]_{i,j} > 0$ and antagonistic if $[{\mathcal A}]_{i,j} <0$. 
Also,  $[ {\mathcal A}]_{i,i} =0,$ for all $i\in [1,N]$. 
We also assume that the  graph ${\mathcal G}$ is  connected and    {\color{black} {\em clustering balanced, with three clusters}, i.e., all the agents are grouped in 3  clusters, ${\mathcal V}_i, i\in [1,3],$ with $n_i=|{\mathcal V}_i|$, such that for every $i,j\in {\mathcal V}_p, p\in [1,k],$ $[{\mathcal A}]_{i,j}\ge 0$, while for every $i\in {\mathcal V}_p, j\in {\mathcal V}_q$, $p,q\in [1,k]$, $p\ne q$, $[{\mathcal A}]_{i,j}\le 0$.
However, the agents cannot be grouped into a smaller number of clusters.} \\
 
 In this paper 
we want to extend some recent results obtained for tripartite consensus of a multi-agent system with 
undirected, signed,  weighted, connected and clustering balanced communication graph,
by relaxing the   homogeneity constraint regarding mutual relationships between agents introduced in \cite{GiuliaElenaAut2020,GiuliaElenaCDC2020}. Specifically, in \cite{GiuliaElenaAut2020,GiuliaElenaCDC2020} we proved that if each agent in a cluster distributes the same amount of ``trust" to the agents in its own group and ``distrust" to the agents belonging to adverse clusters, then it is possible to adopt a slightly modified version of   DeGroot's algorithm in such a way that agents belonging to the same cluster ${\mathcal V}_i, i\in[1,3],$
 asymptotically converge to the same decision, i.e., 
$$\lim_{t\rightarrow +\infty}x_k(t)= c_i, \qquad \forall\ k\in {\mathcal V}_i.$$
We now want to explore under what conditions tripartite consensus can still be achieved even if the aforementioned homogeneity constraint  is removed.

For the sake of simplicity, in the following we will assume that the agents are ordered in such a way that the first $n_1$  agents belong to the  cluster ${\mathcal V}_1$, the subsequent $n_2$ to the cluster ${\mathcal V}_2$, and the   last $n_3$ to the  cluster ${\mathcal V}_3$.  Clearly, $n_1+n_2+n_3=N$. This assumption entails no loss of generality, since it is always possible to reduce ourselves to this structure by means of  a relabelling of the nodes/agents. 
Accordingly, the adjacency matrix of the graph $\mathcal{G}$ can be block-partitioned as follows
\begin{equation}\label{adjacency_m}
{\mathcal  A}=\left[
\begin{array}{ccc}
     {\mathcal A}_{1,1}& {\mathcal A}_{1,2} & {\mathcal A}_{1,3}\\
    
     {\mathcal A}_{2,1}& {\mathcal A}_{2,2} & {\mathcal A}_{2,3}\\
    
     {\mathcal A}_{3,1} & {\mathcal A}_{3,2} & {\mathcal A}_{3,3}
\end{array} \right]
\end{equation}
with ${\mathcal A}_{i,j} \in \mathbb{R}^{n_i \times n_j}$,    ${\mathcal A}_{i,i} = {\mathcal A}_{i,i}^\top \geq 0$,  and
 ${\mathcal A}_{i,j}= {\mathcal A}_{j,i}^\top \leq 0$ $\forall i \neq j$, $i,j \in [1,3]$, $[{\mathcal A}_{i,i}]_{k,k}=0,$ $\forall i \in [1,3], k \in [1,n_i]$.\\
We
consider a distributed control law for the system \eqref{model} of the type
\begin{equation}\label{u}
{\bf u}(t) = -{\mathcal M}{\bf x}(t), 
\end{equation}
where ${\bf {\mathcal{M}}} \in \mathbb{R}^{N\times N}$ takes the form 
\begin{equation}\label{M}
{\bf {\mathcal{M}}} = {\bf  \mathcal{D}-\mathcal{A}}, 
\end{equation}
with 
$\mathcal{D}\in \mathbb{R}^{N\times N}$ a diagonal matrix partitioned according to the block-partition of ${\mathcal A}$, namely
\be
{\mathcal D} = {\rm diag}\{ {\mathcal D}_1, {\mathcal D}_2, {\mathcal D}_3\},
\label{matriceD}
\ee
and 
${\mathcal D}_i \in {\mathbb R}^{n_i}\times {\mathbb R}^{n_i}$, $n_i$ being the cardinality of the $i$th cluster,  $i\in[1,3]$. The diagonal entries  of  ${\mathcal D}_i$ represent the degree of stubbornness of each  agent in ${\mathcal V}_i$. They quantify how much individuals in the $i$th cluster  are convinced of their own opinions.\\
The overall multi-agent system is hence described as
\be
\dot{\bf x}(t) = - {\mathcal M} {\bf x}(t), 
\label{model_final}
\ee
and the aim of this paper is
to investigate if it is possible to choose the matrices ${\mathcal D}_i$   so that 
all the agents reach {\em tripartite consensus}, by this meaning that for almost every      initial condition\footnote{In fact, there is necessarily a subset of ${\mathbb R}^N$, with zero Lebesgue measure, corresponding to which the state trajectory converges to zero.} ${\bf x}(0)\in {\mathbb R}^{N}$ 
all the state variables  associated to agents in the same cluster converge to the same value, namely
\be
\lim_{t\rightarrow +\infty} {\bf x}(t) = [c_1 {\bf 1}_{n_1}^\top,c_2 {\bf 1}_{n_2}^\top,  c_3 {\bf 1}_{n_3}^\top]^\top, 
\label{conditionC3}
\ee for suitable $c_i
= c_i({\bf x}(0)) \in \mathbb{R}, i \in [1,3]$,   not all of them equal to zero.  

{\color{black} It is worth noticing that while  in   the homogeneous case investigated in 
\cite{GiuliaElenaAut2020,GiuliaElenaCDC2020}, {\color{black} the desired goal was achieved  by suitably choosing a stubbornness  degree 
common to all the agents belonging to the same class, now the degree of stubbornness is individually tuned.}

\section{Tripartite consensus: Problem Solution}\label{3}

We first present  necessary and sufficient conditions for tripartite consensus.
\smallskip

 \begin{lemma}\label{lemma1}  \cite{GiuliaElenaAut2020}
Given an undirected, signed,  weighted and connected communication graph,
 $\mathcal{G}$, having  $3$ clusters, the multi-agent system \eqref{model}, with communication graph ${\mathcal G}$ and distributed control law \eqref{u}, and hence described as in \eqref{model_final},
reaches tripartite consensus if and only if the following conditions hold:
\begin{itemize}
    \item[(i)] ${\bf {\mathcal{M}}}$ is a singular positive semi-definite matrix,
    \item[(ii)] The kernel of ${\bf {\mathcal{M}}}$ is spanned by vectors of the type ${\bf v} = [v_1 {\bf 1}_{n_1}^\top,v_2 {\bf 1}_{n_2}^\top,  v_3 {\bf 1}_{n_3}^\top]^\top, v_i \in \mathbb{R}, i \in [1,3]$.
  \end{itemize}
\end{lemma}
\smallskip

We now focus on the previous condition (ii), and provide the following   lemma, whose easy proof is omitted. 
\smallskip

 \begin{lemma}\label{lemmanuovo} 
 Given the matrix  ${\bf {\mathcal{M}}} \in \mathbb{R}^{N\times N}$  described as in
\eqref{M},   $\mathcal{D}\in \mathbb{R}^{N\times N}$ described as in \eqref{matriceD}
and 
${\mathcal D}_i  \in {\mathbb R}^{n_i\times n_i}$, for $i\in [1,3]$, diagonal matrices, 
the kernel of ${\bf {\mathcal{M}}}$ includes a vector of the type ${\bf v} = [v_1 {\bf 1}_{n_1}^\top,v_2 {\bf 1}_{n_2}^\top,  v_3 {\bf 1}_{n_3}^\top]^\top, v_i \in \mathbb{R}, i \in [1,3]$, if and only if 
\be
{\rm rank}   \left( \begin{bmatrix}
   {\bf d}_1 -  {\bf a}_{11} & - {\bf a}_{12} & - {\bf a}_{13} \\
   - {\bf a}_{21} &  {\bf d}_2 -{\bf a}_{22} & -{\bf a}_{23} \\
  -  {\bf a}_{31} &- {\bf a}_{32} &  {\bf d}_3 -{\bf a}_{33}
    \end{bmatrix}\right) < 3,
    \label{rank_cond}
    \ee
 where  
$$
{\bf d}_i := {\mathcal D}_i {\bf 1}_{n_i}, \qquad {\bf a}_{ij} := {\mathcal A}_{i,j}{\bf 1}_{n_j},
\qquad  i,j \in [1,3].$$
\end{lemma}
\smallskip

Based on Lemmas \ref{lemma1} and \ref{lemmanuovo}, in the sequel we will provide conditions ensuring the existence of diagonal matrices
${\mathcal D}_i$ (equivalently, of vectors ${\bf d}_i= {\mathcal D}_i {\bf 1}_{n_i}$), for $i\in [1,3]$, such that the corresponding matrix 
${\bf {\mathcal{M}}}$ is a singular positive semi-definite matrix,
with a simple eigenvalue in $0$ and condition \eqref{rank_cond} holds. In fact,  if $0$   is a simple eigenvalue, in order to fulfil condition (ii) of Lemma \ref{lemma1} 
 it is sufficient to prove that there exists a single vector in the kernel of ${\mathcal M}$ having the desired block structure.
 
It is worth noticing that $ {\bf a}_{ij}  \ne 0$ for every pair $i,j\in [1,3], i\ne j$. In fact ${\bf a}_{ij}=0$ implies ${\mathcal A}_{i,j}=0$ and hence also ${\mathcal A}_{j,i}=0$ which means that ${\mathcal V}_i$ and ${\mathcal V}_j$ could be grouped together, thus contradicting the minimality of the partitioning into $3$ clusters introduced in Assumption 1.
\smallskip

To solve the tripartite consensus problem, we introduce the same ``close friendship" assumption that 
we have adopted in \cite{GiuliaElenaAut2020,GiuliaElenaCDC2020} and that strengthens the relationships among agents in the same cluster.
\smallskip

 {\bf Assumption 2 on the communication structure.}\  [Close friendship]\ 
There exist  two distinct indices   $i_1$ and  $i_2$ in $[1,3]$ such that 
 the cluster ${\mathcal V}_{i_2}$ either consists of a single node/agent or for every pair of distinct agents 
  $(i,j)\in {\mathcal V}_{i_2}\times {\mathcal V}_{i_2}$
either one of the following cases applies  (see Figure \ref{assumption2}): 
\begin{itemize}
\item[i)]    $(i,j)$ are {\em friends} (i.e., the edge $(i,j)$ belongs to ${\mathcal E}$ and it has positive weight); 
\item[ii)]  $(i,j)$  are {\em enemies} (i.e., the edge $(i,j)$ belongs to ${\mathcal E}$ and it has negative weight) of two (not necessarily distinct) vertices  $r$ and $s$ in ${\mathcal V}_{i_1}$
that 
{\color{black} belong to the same connected component in ${\mathcal V}_{i_1}$.}
\end{itemize}
 \begin{figure}[H]
\begin{center}
     \centering
     \includegraphics[scale=0.35]{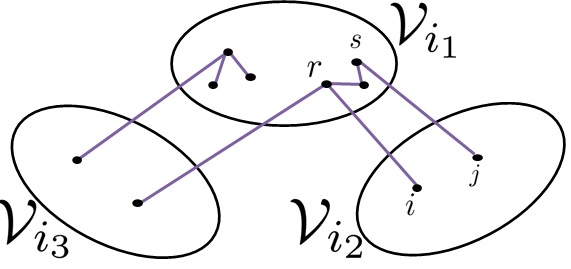}
     \caption{Graphical representation of Assumption 2.}
     \label{assumption2}
\end{center}
 \end{figure}

We are now in a position to introduce one of the two main results of the paper.
\smallskip

\begin{theorem} \label{teo2} 
Consider the multi-agent system \eqref{model}, with undirected, signed,  weighted and connected communication graph
 $\mathcal{G}$ satisfying Assumption 1 and Assumption 2 for a suitable choice of
 $i_1, i_2\in [1,3], i_1\ne i_2.$
 Also, suppose that the following conditions hold:
\begin{itemize}
\item[1)]  every agent in ${\mathcal V}_{i_3}$ has at least one enemy in  ${\mathcal V}_{i_2}$, namely  $\mathcal{A}_{i_3,i_2}{\bf1}_{n_{i_2}}\ll 0$,  and
\item[2)]
  there exists $h\in \{i_2,i_3\}$ such that 
every agent in $\mathcal{V}_{i_1}$  has at least  one enemy in $\mathcal{V}_{h}$, namely
$\mathcal{A}_{i_1,h}{\bf1}_{h}\ll 0$.
\end{itemize}
Then  there exist   diagonal matrices ${\mathcal D}_i \in {\mathbb R}^{n_i\times n_i}, i \in [1,3],$ such that
 the distributed control law \eqref{u}, with    ${\bf {\mathcal{M}}} \in \mathbb{R}^{N\times N}$  described as in
\eqref{M},   $\mathcal{D}\in \mathbb{R}^{N\times N}$ described as in \eqref{matriceD},
makes  the closed-loop multi-agent system \eqref{model_final}
reach tripartite consensus.
\end{theorem}
\smallskip

\begin{proof} 
 We can always relabel the vertices in ${\mathcal V}$ so that   $i_1=1, i_2=3$ and $i_3=2$. Note, also, that    ${\bf a}_{ij} = {\mathcal A}_{i,j}{\bf 1}_{n_j} \le 0$, for every $i,j\in [1,3], i\ne j$ (but ${\bf a}_{ij}\ne 0$), and ${\bf a}_{ii} = {\mathcal A}_{i,i}{\bf 1}_{n_i} \ge 0$.
\\
Based on the previous comments, related to Lemmas \ref{lemma1} and \ref{lemmanuovo}, we prove that there exist diagonal matrices
${\mathcal D}_i$,   $i\in [1,3]$, such that {\bf (A)} ${\bf {\mathcal{M}}}$ is a singular positive semi-definite matrix,
with a simple eigenvalue in $0$, and {\bf (B)} the matrix in \eqref{rank_cond} has a nontrivial vector $\tilde {\bf v} := \begin{bmatrix} v_1 & v_2 & v_3\end{bmatrix}^\top$ in its kernel.  \\
 We first prove that Assumption 2 ensures that {\bf (A)} holds.
To ensure that the matrix
\begin{equation}
{\bf \mathcal{M}}=\left[
\begin{array}{c|cc}
     {\mathcal D}_1-{\mathcal A}_{1,1}& -{\mathcal A}_{1,2} & -{\mathcal A}_{1,3}\\
     \hline
     -{\mathcal A}_{2,1}& {\mathcal D}_2-{\mathcal A}_{2,2} & -{\mathcal A}_{2,3}\\
     -{\mathcal A}_{3,1} & -{\mathcal A}_{3,2} & {\mathcal D}_3-{\mathcal A}_{3,3}
\end{array} \right]
\end{equation}
is positive semidefinite, we    impose  (see \cite{BoydVandenberghe}, page 651)  
that the upper diagonal block is positive definite and its Schur complement is positive semi-definite,
 i.e., that conditions \eqref{first_matr_conN_case2}:
\be\label{first_matr_conN_case2}
    {\mathcal D}_1-{\mathcal A}_{1,1} \succ 0 \ee
    and \eqref{first_matr_con2N_case2} hold.  
\begin{figure*}
\be
\begin{bmatrix}
     {\mathcal D}_2-{\mathcal A}_{2,2}-{\mathcal A}_{2,1}({\mathcal D}_1- {\mathcal A}_{1,1})^{-1}{\mathcal A}_{1,2} & -{\mathcal A}_{2,3}-{\mathcal A}_{2,1} ({\mathcal D}_1- {\mathcal A}_{1,1})^{-1}{\mathcal A}_{1,3}\\
     -{\mathcal A}_{3,2}-{\mathcal A}_{3,1} ({\mathcal D}_1- {\mathcal A}_{1,1})^{-1} {\mathcal A}_{1,2} & {\mathcal D}_3-{\mathcal A}_{3,3}-{\mathcal A}_{3,1} ({\mathcal D}_1- {\mathcal A}_{1,1})^{-1}{\mathcal A}_{1,3}
    \end{bmatrix} \succeq 0.
    \label{first_matr_con2N_case2}
    \ee
\hrulefill
\end{figure*}

\noindent Assume that  \begin{equation}\label{d1_constraint_case2}
 {\bf d}_{1} \gg  {\bf a}_{11} \ge  0.
\end{equation} 
Then $({\mathcal D}_{1} - {\mathcal A}_{1,1}) {\bf 1}_{n_1}\gg 0$, and hence Lemma \ref{lemma3}, part i), holds for 
 ${\bf v} = {\bf 1}_{n_1}$, thus ensuring that ${\mathcal D}_1-{\mathcal A}_{1,1}$ is positive definite.
 
To ensure that  \eqref{first_matr_con2N_case2} holds, we iterate  the same procedure, and impose 
condition: 
\be
{\mathcal D}_{2}-{\mathcal A}_{2,2}-{\mathcal A}_{2,1} ({\mathcal D}_1- {\mathcal A}_{1,1})^{-1}{\mathcal A}_{1,2} \succ 0,
\label{second_matr_conN_case2}
\ee
as well as condition \eqref{second_matr_con2N_case2}.
\begin{figure*}
\begin{eqnarray}
\Phi_3 &:=&{\mathcal D}_3-{\mathcal A}_{3,3}-{\mathcal A}_{3,1} ({\mathcal D}_1- {\mathcal A}_{1,1})^{-1}{\mathcal A}_{1,3}
- [ {\mathcal A}_{3,2}+{\mathcal A}_{3,1} ({\mathcal D}_1- {\mathcal A}_{1,1})^{-1} {\mathcal A}_{1,2}]\nonumber\\
&&\quad \cdot
  [{\mathcal D}_2-{\mathcal A}_{2,2}-{\mathcal A}_{2,1}({\mathcal D}_1- {\mathcal A}_{1,1})^{-1}{\mathcal A}_{1,2}]^{-1}  
[{\mathcal A}_{2,3}+{\mathcal A}_{2,1} ({\mathcal D}_1- {\mathcal A}_{1,1})^{-1}{\mathcal A}_{1,3}] \succeq 0
\ {\rm  and \ singular}.
    \label{second_matr_con2N_case2}
    \end{eqnarray}
 \begin{center}
-------------------------------------------------------------------------------------------------------------------------------------------------
\end{center}
\end{figure*}

To address condition \eqref{second_matr_conN_case2},
we first observe that by Lemma \ref{lemma3}, part ii), $({\mathcal D}_1- {\mathcal A}_{1,1})^{-1}$ is symmetric and nonnegative,  and hence so is 
${\mathcal A}_{2,2}+ {\mathcal A}_{2,1} ({\mathcal D}_1- {\mathcal A}_{1,1})^{-1}{\mathcal A}_{1,2}.$
But then we can apply Lemma \ref{lemma3}, part i), again, by assuming
$D={\mathcal D}_2$ and $A= {\mathcal A}_{2,2}+ {\mathcal A}_{2,1} ({\mathcal D}_1- {\mathcal A}_{1,1})^{-1}{\mathcal A}_{1,2}.$ Indeed, if we   impose 
 the following constraint on ${\bf d}_2$:
\begin{equation}\label{d2_constraint_case2}
    {\bf d}_2\gg  {\bf a}_{22}+ {\mathcal A}_{2,1} ({\mathcal D}_1- {\mathcal A}_{1,1})^{-1} {\bf a}_{12} \ge 0,
\end{equation}
then it is easy to verify that 
\begin{eqnarray*}
(D- A){\bf 1}_{n_2} 
\!\!\!\!&=&\!\!\!\! {\bf d}_2 - {\bf a}_{22} - {\mathcal A}_{2,1} ({\mathcal D}_1- {\mathcal A}_{1,1})^{-1} {\bf a}_{12} \gg 0.
\end{eqnarray*}
 Therefore $D- A$ is positive definite, namely 
\eqref{second_matr_conN_case2} holds. \\
On the other hand, 
we can always choose (see \cite{GiuliaElenaAut2020}) the positive diagonal entries of the diagonal matrix ${\mathcal D}_2$,  namely the vector ${\bf d}_2$, so that not only ${\bf d}_2$ fulfils condition  \eqref{d2_constraint_case2}, but it is also
sufficiently  large to ensure that the entries of 
$[ {\mathcal A}_{3,2}+{\mathcal A}_{3,1} ({\mathcal D}_1- {\mathcal A}_{1,1})^{-1} {\mathcal A}_{1,2}]  [{\mathcal D}_2-{\mathcal A}_{2,2}-{\mathcal A}_{2,1}({\mathcal D}_1- {\mathcal A}_{1,1})^{-1}{\mathcal A}_{1,2}]^{-1}  
[{\mathcal A}_{2,3}+{\mathcal A}_{2,1} ({\mathcal D}_1- {\mathcal A}_{1,1})^{-1}{\mathcal A}_{1,3}]$  are small enough to guarantee that
$$ -(\Phi_3 - {\mathcal D}_3)\approx {\mathcal A}_{3,3}+{\mathcal A}_{3,1} ({\mathcal D}_1- {\mathcal A}_{1,1})^{-1}{\mathcal A}_{1,3}.$$ 
By Assumption 2, for $i_1=1$ and $i_3=2$, the matrix 
$ {\mathcal A}_{3,3}+{\mathcal A}_{3,1} ({\mathcal D}_1- {\mathcal A}_{1,1})^{-1}{\mathcal A}_{1,3}$ has    positive off-diagonal entries, and hence the same is true for $ -(\Phi_3 - {\mathcal D}_3)$. This ensures that $-\Phi_3$ is an irreducible Metzler matrix.

So, now, we are remained with proving that for a suitable choice of ${\mathcal D}_3$ we can ensure that 
\eqref{second_matr_con2N_case2} holds.
If we   apply the vector ${\bf 1}_{n_3}$ on the right side of the matrix $\Phi_3$,
by making use of reasonings similar to those just exploited to prove \eqref{second_matr_conN_case2}, we obtain 
\begin{eqnarray*}
\Phi_3  {\bf 1}_{n_3} &=& {\bf d}_3 
 - {\bf a}_{33}-{\mathcal A}_{3,1} ({\mathcal D}_1- {\mathcal A}_{1,1})^{-1}{\bf a}_{13}\\ 
 &-&
 [ {\mathcal A}_{3,2}+{\mathcal A}_{3,1} ({\mathcal D}_1- {\mathcal A}_{1,1})^{-1} {\mathcal A}_{1,2}] \\
&\cdot& 
  [{\mathcal D}_2-{\mathcal A}_{2,2}-{\mathcal A}_{2,1}({\mathcal D}_1- {\mathcal A}_{1,1})^{-1}{\mathcal A}_{1,2}]^{-1} 
  \\ 
&\cdot&[{\bf a}_{23}+{\mathcal A}_{2,1} ({\mathcal D}_1- {\mathcal A}_{1,1})^{-1} {\bf a}_{13}].
 \end{eqnarray*}
Therefore, by imposing
 \begin{eqnarray}
 {\bf d}_3 
 &=& {\bf a}_{33}+ {\mathcal A}_{3,1} ({\mathcal D}_1- {\mathcal A}_{1,1})^{-1}{\bf a}_{13} \label{d3_constraint_case2} \\ 
 &+&
 [ {\mathcal A}_{3,2}+{\mathcal A}_{3,1} ({\mathcal D}_1- {\mathcal A}_{1,1})^{-1} {\mathcal A}_{1,2}] \nonumber \\
&\cdot& 
  [{\mathcal D}_2-{\mathcal A}_{2,2}-{\mathcal A}_{2,1}({\mathcal D}_1- {\mathcal A}_{1,1})^{-1}{\mathcal A}_{1,2}]^{-1} 
 \nonumber \\ 
&\cdot&[{\bf a}_{23}+{\mathcal A}_{2,1} ({\mathcal D}_1- {\mathcal A}_{1,1})^{-1} {\bf a}_{13}], \nonumber
 \end{eqnarray}
we ensure that 
  $\Phi_3   {\bf 1}_{n_3} = 0.$ 
This guarantees that the matrix $\Phi_3$ has $0$ as an eigenvalue  corresponding to the eigenvector ${\bf1}_{n_3}\gg 0$, and therefore (see \cite{Berman-Plemmons}) $0$ is the  simple dominant eigenvalue of the irreducible Metzler matrix $-\Phi_3$. Since the eigenvalues of ${\bf \mathcal{M}}$ are the union of the eigenvalues of the matrices in \eqref{first_matr_conN_case2} and \eqref{second_matr_conN_case2}  and of the matrix $\Phi_3$, that have been obtained from ${\bf 
\mathcal{M}}$ by applying the Schur complement, then ${\bf 
\mathcal{M}}$ is positive semidefinite with $0$ as simple eigenvalue and thus condition  {\bf (A)} holds.
 Now   we show that under conditions 1) and 2)
 we can determine vectors ${\bf d}_i, i\in [1,3]$, so that also condition {\bf (B)} holds.

Note that by assumption 1), ${\bf a}_{23} \ll 0$, and by assumption 2),  either ${\bf a}_{12} \ll 0$ or $ {\bf a}_{13}\ll 0$.
In the sequel we will focus on the case  ${\bf a}_{12} \ll 0$, the other case being completely equivalent. We want to prove that
we can always find vectors ${\bf d}_i, i\in [1,3]$, consistent with the constraints \eqref{d1_constraint_case2}, \eqref{d2_constraint_case2} and \eqref{d3_constraint_case2}, 
so that {\bf (B)} holds and hence there exist
$v_2, v_3 $ such  that 
\begin{equation}
    \begin{bmatrix}
    {\bf a}_{11} & {\bf a}_{12} & {\bf a}_{13} \\
    {\bf a}_{21} & {\bf a}_{22} & {\bf a}_{23} \\
    {\bf a}_{31} & {\bf a}_{32} & {\bf a}_{33}
    \end{bmatrix}
    \begin{bmatrix}
1 \\
     v_2\\
    v_3
    \end{bmatrix} = 
    \begin{bmatrix}
 {\bf d}_{1}\\
   v_2 {\bf d}_2\\
 v_3 {\bf d}_3
    \end{bmatrix}.
\end{equation}
This is equivalent  to determining scalars  $v_2$ and $v_3$ that make the vectors
    \begin{eqnarray}
        {\bf d}_1 &=& {\bf a}_{11} + v_2 {\bf a}_{12}+ v_3 {\bf a}_{13} \label{d1_eig_case3} \\
        {\bf d}_2 &=& {\bf a}_{22}+ \frac{1}{v_2} {\bf a}_{21}+ \frac{v_3}{v_2} {\bf a}_{23}  \label{d2_eig_case3}\\
        {\bf d}_3 &=& {\bf a}_{33}+ \frac{1}{v_3} {\bf a}_{31} + \frac{v_2}{v_3} {\bf a}_{32} \label{d3_eig_case3},
    \end{eqnarray}
consistent with the    constraints \eqref{d1_constraint_case2}, \eqref{d2_constraint_case2} and \eqref{d3_constraint_case2}.\\
We first note that since ${\bf a}_{12} \ll 0$, we can always choose $v_2 <0$ with large module,  and $v_3>0$ and small, so that
${v_2} {\bf a}_{12}+ v_3 {\bf a}_{13}\gg 0,$ which automatically implies that ${\bf d}_1$ satisfies condition \eqref{d1_constraint_case2}.
Also, we can choose the modules of $v_2$ and $v_3$ in such a way that the entries of ${\bf d}_1$ and hence of ${\mathcal D}_1$ are so large
that ${\bf a}_{23}+ {\mathcal A}_{2,1}(\mathcal{D}_1-\mathcal{A}_{1,1})^{-1} {\bf a}_{13}\approx  {\bf a}_{23} \ll0$ and hence
${\bf a}_{23}+ {\mathcal A}_{2,1}(\mathcal{D}_1-\mathcal{A}_{1,1})^{-1} {\bf a}_{13}  \ll 0,$
and also
\begin{equation}
\frac{v_3}{v_2}\big[{\bf a}_{23}+ {\mathcal A}_{2,1}(\mathcal{D}_1-\mathcal{A}_{1,1})^{-1} {\bf a}_{13} \big] \gg 0.
\label{2nd_con_mod_case3}
\end{equation}
By making use of \eqref{2nd_con_mod_case3}, we obtain that
\begin{eqnarray*}
{\bf d}_2 &=& {\bf a}_{22}+ \frac{1}{v_2} {\bf a}_{21}+ \frac{v_3}{v_2} {\bf a}_{23}\\
&\gg&{\bf a}_{22}+ \frac{1}{v_2} {\bf a}_{21} -  \frac{v_3}{v_2} {\mathcal A}_{2,1}(\mathcal{D}_1-\mathcal{A}_{1,1})^{-1} {\bf a}_{13}\\
&=& {\bf a}_{22}+ \frac{1}{v_2}   {\mathcal A}_{2,1}[{\bf 1}_{n_1} - v_3 (\mathcal{D}_1-\mathcal{A}_{1,1})^{-1} {\bf a}_{13}]\\
&=& {\bf a}_{22}+   \frac{1}{v_2}  {\mathcal A}_{2,1} [v_2(\mathcal{D}_1-\mathcal{A}_{1,1})^{-1} {\bf a}_{12}]\\
&=& {\bf a}_{22}+     {\mathcal A}_{2,1}  (\mathcal{D}_1-\mathcal{A}_{1,1})^{-1} {\bf a}_{12},
\end{eqnarray*}
where we used the fact that 
condition \eqref{d1_eig_case3} is equivalent to
 \begin{equation}\label{1n1_case3}
 {\bf 1}_{n_1} = v_2(\mathcal{D}_1-\mathcal{A}_{1,1})^{-1}{\bf a}_{12}+v_3(\mathcal{D}_1-\mathcal{A}_{1,1})^{-1}{\bf a}_{13}.
\end{equation}
So, this proves that also \eqref{d2_constraint_case2} holds.
Finally, it is possible to prove (details are omitted due to page constraints) that if  the identities \eqref{d1_eig_case3} and \eqref{d2_eig_case3} hold, then  the constraints \eqref{d3_eig_case3} and \eqref{d3_constraint_case2} are equivalent.

 Hence we conclude that there exist suitable choices of ${\bf d}_i, i\in [1,3],$ such that both conditions   {\bf (A)} and {\bf (B)}  are fulfilled and hence the overall multi-agent system reaches tripartite consensus.
\end{proof}
\smallskip

\begin{example}\label{example_alphanot0} 
 Consider the  undirected, signed, weighted, connected  and clustered  communication graph, with three clusters of cardinalities $n_1=5$, $n_2=4$, $n_3=2$, respectively, and adjacency matrix composed of the sub-matrices

{\small
\begin{eqnarray*}
\mathcal{A}_{1,1}  \!\!\!\!&=& \!\! \!\!
\begin{bmatrix} 
0& 4& 0& 0& 1\\
4&  0&  3&  10&  2\\
0& 3& 0& 1& 0\\
0& 10& 1& 0& 1\\
1& 2& 0& 1& 0
 \end{bmatrix}\ 
\mathcal{A}_{1,2} =
-\begin{bmatrix}
 1.5& 1.5& 0& 1.5\\
       0.5& 3& 0& 2.5\\
       3& 0.5& 0& 2.5\\
       3& 3& 0& 2\\
       0& 0& 0.5& 2.5
 \end{bmatrix},\\
 \mathcal{A}_{1,3}   \!\!\!\!&=& \!\! \!\! -\begin{bmatrix}7& 2\\
       0& 3\\
       4& 2\\
       8& 0\\
       7& 3\end{bmatrix},\ \mathcal{A}_{2,2} = \begin{bmatrix} 
0& 6& 0& 2\\
       6& 0& 4& 4\\
       0& 4& 0& 0\\
       2& 4& 0& 0
\end{bmatrix},
\\
\mathcal{A}_{2,3}   \!\!\!\!&=&  \!\!\!\! -\begin{bmatrix}3& 4\\
       6& 1\\
       1& 0\\
       4& 6\end{bmatrix}, \ \mathcal{A}_{3,3}= \begin{bmatrix}0&6\\6&0 \end{bmatrix},
\end{eqnarray*}
 }

\noindent all the others being deduced by symmetry. Assumption 2 holds for $i_1=1$ and $i_2=3$, and both assumptions 1) and 2) of Theorem \ref{teo2} hold, since ${\bf a}_{23}\ll 0$, ${\bf a}_{12}$ and ${\bf a}_{13}$ are both strictly negative vectors. So, we can assume, for
example, $(v_1,v_2, v_3) = (1,5,-8)$, and hence
${\bf d}_1= [ 54.5\ 13 \ 19.5 \ 36 \ 74]^\top$, ${\bf d}_2= [ 17.6 \ 23.6 \ 5.5\  19.9]^\top$  and ${\bf d}_3= [18\ 14.125]^\top$. 
The dynamics of the state vector of  the  system, with random initial conditions $\bf{x}(0)$ taken as realizations of a gaussian vector with $0$ mean and variance  $\sigma^{2}=4$, i.e. ${\bf x}(0) \sim \mathcal{N}(0,4)$, is given in Fig. \ref{no_hom_alpha2not0}. The plot shows that tripartite consensus is reached after about 1.8 units of time with regime values{\color{black} $(c_1\ c_2 \ c_3) = (0.22 \ \ 1.11 \  \ -1.76)= 0.22 \cdot (1 \ 5 \ -8)= 0.22\cdot  (v_1 \ v_2 \ v_3)$.}
\begin{figure}[H]
     \centering
     \includegraphics[scale=0.25]{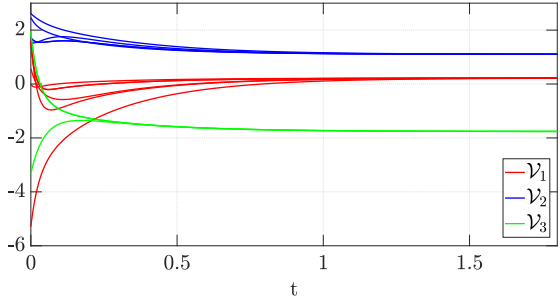}
     \caption{Tripartite consensus for Example 1}
     \label{no_hom_alpha2not0}
\end{figure}
\end{example}
\smallskip

\section{Sign Consensus}\label{4}

In this section we introduce the concept of sign consensus for which a formal definition is given in the following.
\smallskip

\begin{definition}[Sign Consensus] 
The overall multi-agent system described as in \eqref{model_final},   with  ${\bf {\mathcal{M}}} \in \mathbb{R}^{N\times N}$  described as in
\eqref{M},   $\mathcal{D}\in \mathbb{R}^{N\times N}$ described as in \eqref{matriceD}
and 
${\mathcal D}_i  \in {\mathbb R}^{n_i\times n_i}$, for $i\in [1,3]$, diagonal matrices, whose interconnection topology is described by an undirected, signed and connected communication graph $\mathcal{G}$, having 3 clusters, reaches \textit{sign consensus} if  
 there exists a relabelling of the three clusters such that, for every index $i\in {\mathcal V}_2$, 
$\lim_{t\rightarrow \infty} x_i(t) =0$, while for every $i,j\in {\mathcal V}_1\cup {\mathcal V}_3$
\begin{align*}
&\lim_{t\rightarrow \infty}{\rm sgn}(x_i(t))-{\rm sgn}(x_j(t)) = 0, \quad {\rm if}\ \exists m:  i,j \in \mathcal{V}_{m}, \notag\\
&\lim_{t\rightarrow \infty}{\rm sgn}(x_i(t))-{\rm sgn}(x_j(t)) \neq 0, \quad   {\rm if}\ \not\exists m:  i,j \in \mathcal{V}_{m}.\end{align*}
\end{definition}
 \smallskip
 
The following lemma  provides   necessary and sufficient conditions for sign consensus to be reached.
\smallskip

 \begin{lemma}\label{lemma_trip_sign} 
Given an undirected, signed,  weighted and connected communication graph,
 $\mathcal{G}$, having  $3$ clusters, the multi-agent system \eqref{model}, with communication graph ${\mathcal G}$ and distributed control law \eqref{u}, and hence described as in \eqref{model_final}, with    ${\bf {\mathcal{M}}} \in \mathbb{R}^{N\times N}$  given in
\eqref{M},   ${\mathcal A}$    in \eqref{adjacency_m}, $\mathcal{D}\in \mathbb{R}^{N\times N}$  in \eqref{matriceD}
and 
 ${\mathcal D}_i  \in {\mathbb R}^{n_i\times n_i}$, for $i\in [1,3],$ diagonal matrices
reaches sign consensus if and only if the following conditions hold:
\begin{itemize}
    \item[i)] ${\bf {\mathcal{M}}}$ is a singular positive semi-definite matrix.
    \item[ii)]   There exists a reordering $\{i_1,i_2,i_3\}$ of the index set $\{1,2,3\}$ such that every nonzero vector
    in the kernel of ${\bf {\mathcal{M}}}$ can be expressed as ${\bf v} = [{\bf v}_1^\top, {\bf v}_2^\top,  {\bf v}_3^\top]^\top$ 
     with ${\bf v}_{i_2}=0$, and in the pair $({\bf v}_{i_1},{\bf v}_{i_3})$ one of the vectors is strictly positive and one is strictly negative.
  \end{itemize}
\end{lemma}
\smallskip

\begin{proof}
Analogous to the proof of  Lemma \ref{lemma1} (see \cite{GiuliaElenaAut2020}). 
\end{proof}
\smallskip

\begin{theorem} \label{teo_signconsensus}
Consider the multi-agent system \eqref{model}, with undirected, signed,  weighted and connected communication graph
 $\mathcal{G}$ satisfying Assumption 1 and  Assumption 2 for a suitable choice of
 $i_1, i_2\in [1,3], i_1\ne i_2.$
  Also, suppose that  the following conditions hold:
 \begin{itemize}
\item[a)]   every agent in ${\mathcal V}_{i_1}$ has at least one enemy in $\mathcal{V}_{i_2}$,   which means that ${\mathcal A}_{i_1,i_2}{\bf 1}_{n_{i_2}} \ll 0$, and
\item[b)]  there exist vectors ${\bf v}_{i_1} \in \mathbb{R}^{n_{i_1}}$ and  ${\bf v}_{i_2}\in   \mathbb{R}^{n_{i_2}}$,   one of them strictly positive and the other  strictly negative, such that
 $\mathcal{A}_{i_3,i_1}{\bf v}_{i_1} + \mathcal{A}_{i_3,i_2}{\bf v}_{i_2} =0$, where   $i_3 = [1,3]\setminus \{ i_1, i_2 \}$.
 \end{itemize}
Then  there exist   diagonal matrices ${\mathcal D}_i \in {\mathbb R}^{n_i\times n_i}, i \in [1,3],$ such that
 the distributed control law \eqref{u}, with    ${\bf {\mathcal{M}}} \in \mathbb{R}^{N\times N}$  described as in
\eqref{M},   $\mathcal{D}\in \mathbb{R}^{N\times N}$ described as in \eqref{matriceD},
makes  the closed-loop multi-agent system \eqref{model_final}
reach sign consensus.
\end{theorem}
\smallskip

\begin{proof} 
 We can always relabel the vertices in ${\mathcal V}$ so that assumption a)  and b) hold for $i_1=1, i_2=3, i_3=2$.\\
By   Lemma \ref{lemma_trip_sign}, it will be sufficient 
to prove that under assumptions  a)-b), it is  always possible to choose the diagonal matrices
 ${\mathcal D}_1, {\mathcal D}_2$ and ${\mathcal D}_3$ so that {\bf (A)} the matrix ${\bf {\mathcal{M}}}$  is  singular and positive semi-definite with a simple eigenvalue in $0$, and 
{\bf (B)} the
  kernel   of ${\mathcal  M}$  includes the vector   ${\bf v} = [{\bf v}_{1}^\top,{\bf 0}_{n_2}^\top, {\bf v}_{3}^\top]^\top$, where ${\bf v}_{1} \in \mathbb{R}^{n_1}$ and ${\bf v}_{3} \in \mathbb{R}^{n_3}$ are two vectors satisfying assumptions b), and we assume w.lo.g.  that ${\bf v}_1 \gg 0$  and ${\bf v}_3 \ll 0$. 
  \\
To prove {\bf (B)} we note  that solving the system of equations ${\bf \mathcal{M}}{\bf v}=0 {\bf v}$ is equivalent to solve the system
\begin{equation}
    \begin{bmatrix}
    {\mathcal D}_1-{\mathcal A}_{1,1} & -{\mathcal A}_{1,2} & -{\mathcal A}_{1,3} \\
    -{\mathcal A}_{2,1} & {\mathcal D}_2-{\mathcal A}_{2,2} & -{\mathcal A}_{2,3} \\
    -{\mathcal A}_{3,1} & {\mathcal A}_{3,2} & {\mathcal D}_3-{\mathcal A}_{3,3}
    \end{bmatrix}
    \begin{bmatrix}
{\bf v}_1 \\
     {\bf 0}_{n_2}\\
   {\bf v}_{3}
    \end{bmatrix} = {\bf 0}_{N},
\end{equation}
and this in turn is equivalent to the three identities
    \begin{eqnarray}
         {\mathcal D}_1 {\bf v}_1 &=& {\mathcal A}_{1,1}{\bf v}_1 +  {\mathcal A}_{1,3} {\bf v}_{3} \label{d1_eig_sign} \\
        {\bf 0}_{n_2} &=&    {\mathcal A}_{2,1}{\bf v}_1 +   {\mathcal A}_{2,3}{\bf v}_{3}  \label{d2_eig_sign}.\\
        {\mathcal D}_3 {\bf v}_3  &=&  {\mathcal A}_{3,1}{\bf v}_1 + {\mathcal A}_{3,3}{\bf v}_{3}  \label{d3_eig_sign}.
    \end{eqnarray}
    Identity \eqref{d2_eig_sign}  holds by   assumption b)
  and we  note that the constraint \eqref{d1_eig_sign} and \eqref{d3_eig_sign}  allow to uniquely determine 
 \footnote{ Note, however, that the values of ${\mathcal D}_1$ and ${\mathcal D}_3$ depend on the specific choice of the vectors ${\bf v}_1$ and ${\bf v}_3$ satisfying \eqref{d2_eig_sign}, which are not necessarily uniquely determined.} 
  the diagonal matrices ${\mathcal D}_1$ and ${\mathcal D}_3$, since they can be component-wise written as 
\begin{equation}\label{d1d3}
[\mathcal{D}_p]_{i,i} = \frac{1}{[{\bf v}_p]_i}\Bigg(\sum_{j;j\neq i}[{\mathcal A}_{p,p}]_{ij}[{\bf v}_p]_j+\sum_{k=1}^{n_q} [\mathcal{A}_{p,q}]_{i,k}[{\bf v}_{q}]_k\Bigg)
\end{equation} 
for $p,q\in\{1,3\}, p\neq q $.\\
We are now remained with proving that, after having determined the matrices ${\mathcal D}_1$ and ${\mathcal D}_3$, it is always possible to choose ${\mathcal D}_2$ so that   {\bf (A)} is satisfied. To do so 
we    proceed as follows  (see \cite{BoydVandenberghe}, page 651): we first verify 
that the upper diagonal block of ${\mathcal M}$:
 \begin{equation}
{\bf \mathcal{M}}=\left[
\begin{array}{c|cc}
     {\mathcal D}_1-{\mathcal A}_{1,1}& -{\mathcal A}_{1,2} & -{\mathcal A}_{1,3}\\
     \hline
     -{\mathcal A}_{2,1}& {\mathcal D}_2-{\mathcal A}_{2,2} & -{\mathcal A}_{2,3}\\
     -{\mathcal A}_{3,1} & -{\mathcal A}_{3,2} & {\mathcal D}_3-{\mathcal A}_{3,3}
\end{array} \right], 
\end{equation}
is positive definite, namely condition
\be\label{first_matr_conN}
    {\mathcal D}_1-{\mathcal A}_{1,1} \succ 0 \ee
holds, and then impose (by means of a suitable choice of ${\mathcal D}_2$) that its Schur complement is positive semi-definite with a simple eigenvalue in $0$, namely it
 verifies condition
    \eqref{first_matr_con2N}, and it has a simple eigenvalue in $0$.
\begin{figure*}
\be
\begin{bmatrix}
     {\mathcal D}_2-{\mathcal A}_{2,2}-{\mathcal A}_{2,1}({\mathcal D}_1- {\mathcal A}_{1,1})^{-1}{\mathcal A}_{1,2} & -{\mathcal A}_{2,3}-{\mathcal A}_{2,1} ({\mathcal D}_1- {\mathcal A}_{1,1})^{-1}{\mathcal A}_{1,3}\\
     -{\mathcal A}_{3,2}-{\mathcal A}_{3,1} ({\mathcal D}_1- {\mathcal A}_{1,1})^{-1} {\mathcal A}_{1,2} & {\mathcal D}_3-{\mathcal A}_{3,3}-{\mathcal A}_{3,1} ({\mathcal D}_1- {\mathcal A}_{1,1})^{-1}{\mathcal A}_{1,3}
    \end{bmatrix} \succeq 0.
    \label{first_matr_con2N}
    \ee
 \begin{center}
-------------------------------------------------------------------------------------------------------------------------------------------------
\end{center}
\end{figure*}
\noindent   Condition \eqref{d1_eig_sign} ensures that
  \begin{equation}\label{d1_constraint}
 (\mathcal{D}_1 - \mathcal{A}_{1,1}){\bf v}_{1} =  \mathcal{A}_{1,3}{\bf v}_{3} \gg  0,
\end{equation} 
where we used the fact that ${\bf v}_3 \ll 0$ and ${\mathcal A}_{1,3}$ has no zero rows.
Then
Lemma \ref{lemma3}, part i), holds with
 ${\bf v} = {\bf v}_{1}$, thus ensuring that ${\mathcal D}_1-{\mathcal A}_{1,1}$ is positive definite.
 
To ensure that  \eqref{first_matr_con2N} holds  for a suitable choice of ${\mathcal D}_2$, we iterate  the same procedure, and impose 
condition \eqref{second_matr_conN}:
\be
{\mathcal D}_{2,2}-{\mathcal A}_{2,2}-{\mathcal A}_{2,1} ({\mathcal D}_1- {\mathcal A}_{1,1})^{-1}{\mathcal A}_{1,2} \succ 0,
\label{second_matr_conN}
\ee
as well as condition \eqref{second_matr_con2N}.
\begin{figure*}
\begin{eqnarray}
\Phi_3 &:=&{\mathcal D}_3-{\mathcal A}_{3,3}-{\mathcal A}_{3,1} ({\mathcal D}_1- {\mathcal A}_{1,1})^{-1}{\mathcal A}_{1,3}
- [ {\mathcal A}_{3,2}+{\mathcal A}_{3,1} ({\mathcal D}_1- {\mathcal A}_{1,1})^{-1} {\mathcal A}_{1,2}]\nonumber\\
&&\quad \cdot
  [{\mathcal D}_2-{\mathcal A}_{2,2}-{\mathcal A}_{2,1}({\mathcal D}_1- {\mathcal A}_{1,1})^{-1}{\mathcal A}_{1,2}]^{-1}  
[{\mathcal A}_{2,3}+{\mathcal A}_{2,1} ({\mathcal D}_1- {\mathcal A}_{1,1})^{-1}{\mathcal A}_{1,3}] \succeq 0.
    \label{second_matr_con2N}
    \end{eqnarray}
 \begin{center}
-------------------------------------------------------------------------------------------------------------------------------------------------
\end{center}
\end{figure*}

To address condition \eqref{second_matr_conN},
we first observe that by Lemma \ref{lemma3}, part ii), $({\mathcal D}_1- {\mathcal A}_{1,1})^{-1}$ is symmetric and nonnegative,  and hence so is 
$A := {\mathcal A}_{2,2}+ {\mathcal A}_{2,1} ({\mathcal D}_1- {\mathcal A}_{1,1})^{-1}{\mathcal A}_{1,2}.$
 Let us  set
${\bf a}_{i2} := {\mathcal A}_{i,2}{\bf 1}_{n_2},  i\in [1,2],$ and
${\bf d}_2 := {\mathcal D}_2 {\bf 1}_{n_2},$
and  impose the following constraint on ${\bf d}_2$:
\begin{equation}\label{d2_constraint}
    {\bf d}_2\gg  {\bf a}_{22}+ {\mathcal A}_{2,1} ({\mathcal D}_1- {\mathcal A}_{1,1})^{-1} {\bf a}_{12}.
\end{equation}
Then it is easy to verify that 
\begin{eqnarray*}
(D- A){\bf 1}_{n_2} 
\!\!\!\!&=&\!\!\!\! {\bf d}_2 - {\bf a}_{22} - {\mathcal A}_{2,1} ({\mathcal D}_1- {\mathcal A}_{1,1})^{-1} {\bf a}_{12} \gg 0,
\end{eqnarray*}
where $D ={\mathcal D}_2$.
But then we can apply Lemma \ref{lemma3}, part i), again, 
to claim  that $D- A$ is positive definite, namely 
\eqref{second_matr_conN} holds.\\
We now observe (see \cite{GiuliaElenaAut2020}) that we can always choose the positive diagonal entries of the diagonal matrix ${\mathcal D}_2$ sufficiently  large to ensure that   (not only \eqref{d2_constraint} holds, but also) the entries of 
$[ {\mathcal A}_{3,2}+{\mathcal A}_{3,1} ({\mathcal D}_1- {\mathcal A}_{1,1})^{-1} {\mathcal A}_{1,2}]  [{\mathcal D}_2-{\mathcal A}_{2,2}-{\mathcal A}_{2,1}({\mathcal D}_1- {\mathcal A}_{1,1})^{-1}{\mathcal A}_{1,2}]^{-1}  
[{\mathcal A}_{2,3}+{\mathcal A}_{2,1} ({\mathcal D}_1- {\mathcal A}_{1,1})^{-1}{\mathcal A}_{1,3}]$ are arbitrarily small and hence
also the matrix $A= -\Phi_3 + {\mathcal D}_3$ has    positive off-diagonal entries. This ensures that $-\Phi_3$ is an irreducible Metzler matrix.

So,  we now prove that 
\eqref{second_matr_con2N} holds.
We   observe that from   assumption a) for $i_1=1$ and $i_2=3$
it follows that ${\mathcal A}_{3,3}+{\mathcal A}_{3,1} ({\mathcal D}_1- {\mathcal A}_{1,1})^{-1}{\mathcal A}_{1,3}$ is a nonnegative matrix whose off-diagonal entries are all positive.
If we   apply the vector $-{\bf v}_{3}\gg 0$ on the right side of the matrix $\Phi_3$, appearing in \eqref{second_matr_con2N}, 
we obtain 
\begin{eqnarray*}
-\Phi_3  {\bf v}_{3} &=& -\mathcal{D}_3{\bf v}_3 
 +\mathcal{A}_{3,3}{\bf v}_{3}+{\mathcal A}_{3,1} ({\mathcal D}_1- {\mathcal A}_{1,1})^{-1}\mathcal{A}_{1,3}{\bf v}_{3}\\ 
 &+&
 [ {\mathcal A}_{3,2}+{\mathcal A}_{3,1} ({\mathcal D}_1- {\mathcal A}_{1,1})^{-1} {\mathcal A}_{1,2}] \\
&\cdot& 
  [{\mathcal D}_2-{\mathcal A}_{2,2}-{\mathcal A}_{2,1}({\mathcal D}_1- {\mathcal A}_{1,1})^{-1}{\mathcal A}_{1,2}]^{-1} 
  \\ 
&\cdot&[\mathcal{A}_{2,3}{\bf v}_{3}+{\mathcal A}_{2,1} ({\mathcal D}_1- {\mathcal A}_{1,1})^{-1}\mathcal{A}_{1,3} {\bf v}_{3}].
 \end{eqnarray*}
We first note that  by \eqref{d3_eig_sign} we have $- {\mathcal D}_3 {\bf v}_3  + {\mathcal A}_{3,3}{\bf v}_{3}  =  - {\mathcal A}_{3,1}{\bf v}_1.$
On the other hand, from equation \eqref{d1_eig_sign} one gets  
\be
\mathcal{A}_{1,3}{\bf v}_{3} = (\mathcal{D}_1-\mathcal{A}_{1,1}){\bf v}_1,
\label{012}
\ee
from which it follows  
\be
\mathcal{A}_{3,1} (\mathcal{D}_1-\mathcal{A}_{1,1})^{-1}\mathcal{A}_{1,3}{\bf v}_{3} = \mathcal{A}_{3,1}{\bf v}_1.
\label{123}
\ee
Therefore
\be
-\mathcal{D}_3{\bf v}_3 
 +\mathcal{A}_{3,3}{\bf v}_{3}+{\mathcal A}_{3,1} ({\mathcal D}_1- {\mathcal A}_{1,1})^{-1}\mathcal{A}_{1,3}{\bf v}_{3} =0.
 \label{345}
 \ee
 On the other hand, from \eqref{012} it also follows that
   $\mathcal{A}_{2,1} (\mathcal{D}_1-\mathcal{A}_{1,1})^{-1}\mathcal{A}_{1,3}{\bf v}_{3} = {\mathcal A}_{21}{\bf v}_1$,
and making use of \eqref{d2_eig_sign}, this latter identity leads to
   \be
   \mathcal{A}_{2,1} (\mathcal{D}_1-\mathcal{A}_{1,1})^{-1}\mathcal{A}_{1,3}{\bf v}_{3}  + \mathcal{A}_{2,3}{\bf v}_{3}  = 0.
   \label{234}
   \ee
   So, by replacing \eqref{345} and \eqref{234} in the expression of $-\Phi_3{\bf v}_3$ we obtain the zero vector.\\
Since $-\Phi_3$ is an irreducible Metzler matrix, this ensures   \cite{Berman-Plemmons} that $0$ is the  simple dominant eigenvalue of    $-\Phi_3$ and hence $\Phi_3$ is positive semidefinite and singular with a simple eigenvalue in $0$. Since the eigenvalues of ${\bf \mathcal{M}}$ are the union of the eigenvalues of the matrices in \eqref{first_matr_conN} and \eqref{second_matr_conN} and of the matrix $\Phi_3$, that have been obtained from ${\bf 
\mathcal{M}}$ by applying the Schur complement, then ${\bf 
\mathcal{M}}$ is positive semidefinite with a simple eigenvalue in $0$, and hence {\bf (A)} holds.
\\
To conclude, we have proved that by setting ${\mathcal D}_1$ and ${\mathcal D}_3$ as in \eqref{d1d3}, by choosing  
the diagonal entries of the diagonal matrix ${\mathcal D}_2$ sufficiently large,
 both conditions of Lemma \ref{lemma_trip_sign} are fulfilled, and the overall multi-agent system reaches sign consensus.
\end{proof}
\smallskip

\normalsize
\begin{example}
 Consider the  undirected, signed, unweighted, connected  and clustered  communication graph, with three  clusters of cardinality $n_1=5$, $n_2=4$, $n_3=2$, respectively, and adjacency matrix  whose submtarices ${\mathcal A}_{i,i}, i\in [1,3]$ are as in Example 1, while the remaining
 blocks are
 \begin{eqnarray*}
\mathcal{A}_{1,2} &=&
-\begin{bmatrix}
 3& 3& 0& 3\\
       6& 6& 0& 12\\
       6& 6& 0& 6\\
       6& 6& 0& 6\\
       0& 0& 3& 3
 \end{bmatrix},\
 \mathcal{A}_{1,3} = -\begin{bmatrix}3.5& 1\\
       0& 1.5\\
       1& 1\\
       1.5& 0\\
       3.5& 1.5\end{bmatrix},\\
\mathcal{A}_{2,3} &=&-\begin{bmatrix}36& 3\\
       24& 6\\
       12& 0\\
       12& 15\end{bmatrix}.
\end{eqnarray*}
Condition b) of Theorem \ref{teo_signconsensus} holds for   ${\bf v}_1 = [2,1,1,1,2]^\top\gg 0$, ${\bf v}_3 = -[0.5,2]^\top\ll 0$, while ${\bf d}_2={\bf a}_{22}+\mathcal{A}_{2,1}(\mathcal{D}_1-\mathcal{A}_{1,1})^{-1}{\bf a}_{12}+\bf{v}$, where $\bf{v}$ is a vector whose entries are the absolute value of the entries of the realization of a gaussian vector with $0$ mean and standard deviation $\sigma=200$. 

The dynamics of the state vector of    system   \eqref{model_final}, with random initial conditions $\bf{x}(0)$ taken as realizations of a gaussian vector with $0$ mean and variance  $\sigma^{2}=4$, i.e. ${\bf x}(0) \sim \mathcal{N}(0,4)$, is given in Fig. \ref{sign_con}.
 \begin{figure}[H]
     \centering
     \includegraphics[scale=0.25]{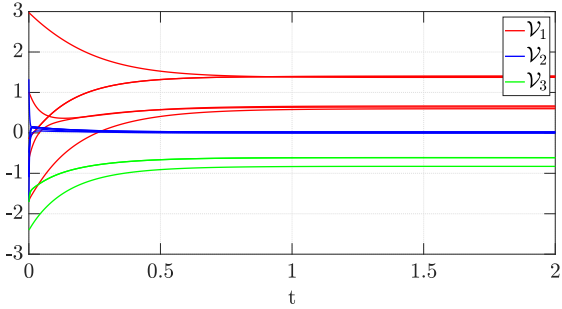}
     \caption{Sign consensus for Example 2.}
     \label{sign_con}
\end{figure}
  \end{example}

\section{Conclusions}
 
{\color{black} In this paper we have investigated the tripartite  and the sign consensus problems for a multi-agent system whose agents are described as simple integrators, and whose communications graph is undirected, signed and weighted, and clustered into three groups. 
 By extending the homogeneous case addressed in \cite{GiuliaElenaAut2020, GiuliaElenaCDC2020}, we have been able to show that, under some assumptions on the structure of the  graph (and, in the case of sign consensus, also on the weights of some edges)
 a modified version of the DeGroot's algorithm, with individually tuned stubbornness degrees, allows to achieve the desired goal.
 \\
 The design of the proposed algorithm is centralised, but its implementation is distributed.
 Also, the algorithm we propose is not adaptive and in general if the weights change, the feedback control law needs to be re-tuned. An exception is represented by the case when, even if the topology changes, the partition into clusters does not and for every agent $i$ the sum of the weights of all the incoming edges  $(j,i)$, where $j$ is an agent of some cluster ${\mathcal V}_k$, does not change  for every $k\in [1,3]$. This means that the vectors ${\bf a}_{ij}(t) = {\mathcal A}_{i,j}(t) {\bf 1}_{n_j}$ are time invariant even if the entries of the   matrices ${\mathcal A}_{i,j}(t)$ change values (but   not  signs).
 
 Future work will try to address the general   case of $k \ge 3$ clusters,
 to weaken the constraints on the communication graph and to explore whether it is possible to devise some distributed algorithm
 that allows agents to update  their own stubbornness degrees to obtain the desired  form of consensus.}

\bibliographystyle{plain}
\bibliography{Refer158}

\begin{thebibliography}{10}

\bibitem{Altafini2013}
C.~Altafini.
\newblock Consensus problems on networks with antagonistic interactions.
\newblock {\em IEEE Trans. Aut. Contr.}, 58, no. 4:935--946, 2013.

\bibitem{Bauso1}
D.~Bauso, L.~Giarr\`e, and R.~Pesenti.
\newblock Quantized dissensus in networks of agents subject to death and
  duplication.
\newblock {\em IEEE Trans. Aut. Contr}, 57:783--788, 2012.

\bibitem{Berman-Plemmons}
A.~Berman and R.J. Plemmons.
\newblock {\em Nonnegative matrices in the mathematical sciences}.
\newblock Academic Press, New York, 1979.

\bibitem{BoydVandenberghe}
S.~Boyd and L.~Vandenberghe.
\newblock {\em Convex Optimization}.
\newblock Cambridge University Press, 2004.

\bibitem{Bullo2020}
P.~Cisneros-Velarde and F.~Bullo.
\newblock Signed network formation games and clustering balance.
\newblock {\em Dynamic Games and Applications}, 2020.

\bibitem{davis67}
J.A. Davis.
\newblock Clustering and structural balance in graphs.
\newblock {\em SAGE Social Science Collections}, 20(2):181--187, 1957.

\bibitem{DeGroot}
M.H. DeGroot.
\newblock Reaching a consensus.
\newblock {\em J. Amer. Statist. Assoc.}, 69, no. 345:118--121, 1974.

\bibitem{Easley}
D.~Easley and J.~Kleinberg.
\newblock {\em Networks, Crowds, and Markets. Reasoning About a Highly
  Connected World}.
\newblock Cambridge Univ. Press, Cambridge, U.K., 2010.

\bibitem{BookFarina}
L.~Farina and S.~Rinaldi.
\newblock {\em Positive linear systems: theory and applications}.
\newblock Wiley-Interscience, Series on Pure and Applied Mathematics, New York,
  2000.

\bibitem{F12b}
G.F. Frobenius.
\newblock {\"U}ber {M}atrizen aus nicht {N}egativen {E}lementen.
\newblock {\em Sitzungsberichte der K{\"o}niglich Preussischen Akademie der
  Wissenschaften, Berlin, Germany, 1912, pp.456-477}, reprinted in Ges. Abh.,
  Springer, Berlin, Germany, vol.3:546--567, 1968.

\bibitem{Heider}
F.~Heider.
\newblock Social perception and phenomenal causality.
\newblock {\em Psycological Review}, 51(6):358--374, 1944.

\bibitem{Jiang}
T.~Jiang and J.~Baras.
\newblock Trust evaluation in anarchy: a case study on autonomous networks.
\newblock pages 23--29, Barcelona, Spain, 2006.

\bibitem{Johnsen}
E.C. Johnsen.
\newblock The micro-macro connection: exact structure and process.
\newblock In Roberts F., editor, {\em Applications of Combinatorics and Graph
  Theory to the Biological and Social Sciences. The IMA Volumes in Mathematics
  and Its Appl.}, volume~17, pages 169--201. Springer,New York, NY, 1989.

\bibitem{Minc}
H.~Minc.
\newblock {\em Nonnegative Matrices}.
\newblock J.Wiley \& Sons, New York, 1988.

\bibitem{OlfatiFaxMurray}
R.~Olfati-Saber, J.A. Fax, and R.M. Murray.
\newblock Consensus and cooperation in networked multi-agent systems.
\newblock {\em Proc. of the IEEE}, 95, no. 1:215--233, 2007.

\bibitem{OF-Murray2004}
R.~Olfati-Saber and R.M. Murray.
\newblock Consensus problems in networks of agents with switching topology and
  time-delays.
\newblock {\em IEEE Trans. Aut. Contr.}, 49, no. 9:1520 --1533, 2004.

\bibitem{GiuliaElenaAut2020}
G.~De Pasquale and M.E. Valcher.
\newblock Consensus for clusters of agents with cooperative and antagonistic
  relationships.
\newblock {\em submitted, available at http://arxiv.org/abs/2008.12398}, 2020.

\bibitem{GiuliaElenaCDC2020}
G.~De Pasquale and M.E. Valcher.
\newblock Consensus problems on clustered networks.
\newblock In {\em Proc. of the 59th IEEE Conf. Decision and Control}, pages
  3675--3680, Jeju Island, Republic of Korea, 2020.

\bibitem{ProskurnikovTempo}
A.V. Proskurnikov and R.~Tempo.
\newblock A tutorial on modeling and analysis of dynamic social networks. part
  i.
\newblock {\em Annu. Rev. Control.}, 43:65--79, 2017.

\bibitem{QinMaZhengGao}
J.~Qin, Q.~Ma, W.X. Zheng, and H.~Gao.
\newblock $h_\infty$ group consensus for clusters of agents with model
  uncertainty and external disturbance.
\newblock In {\em Prof. of the 54th IEEE Conference on Decision and Control},
  pages 2841--2846, Osaka, Japan, 2015.

\bibitem{QinYu2013}
J.~Qin and C.~Yu.
\newblock Cluster consensus control of generic linear multi-agent systems under
  directed topology with acyclic partition.
\newblock {\em Automatica}, 49(9):2898--2905, 2013.

\bibitem{QinYuAnderson2016}
J.~Qin, C.~Yu, and B.D.O. Anderson.
\newblock On leaderless and leader-following consensus for interacting clusters
  of second-order multi-agent systems.
\newblock {\em Automatica}, 74:214--221, 2016.

\bibitem{RenBeardAtkins}
W.~Ren, R.W. Beard, and E.M. Atkins.
\newblock Information consensus in multivehicle cooperative control.
\newblock {\em IEEE Control Sys. Magazine}, 27 (2):71--82, 2007.

\bibitem{SonHinrichsen}
N.K. Son and D.~Hinrichsen.
\newblock Robust stability of positive continuous time systems.
\newblock {\em Numerical Functional Analysis and Optimization}, 17 (5 \&
  6):649--659, 1996.

\bibitem{ValcherMisra}
M.E. Valcher and P.~Misra.
\newblock On the consensus and bipartite consensus in high-order multi-agent
  dynamical systems with antagonistic interactions.
\newblock {\em Systems \& Control Letters}, 66(1):94--103, 2014.

\bibitem{XiaCao2011}
W.~Xia and M.~Cao.
\newblock Clustering in diffusively coupled networks.
\newblock {\em Automatica}, 47(11):2395--2405, 2011.

\bibitem{YuWang2010}
J.~Yu and L.~Wang.
\newblock Group consensus in multi-agent systems with switching topologies and
  communication delays.
\newblock {\em Systems and Control Letters}, 59(6):340--348, 2010.

\end{thebibliography}
\end{document}